# ON THE RELAXATION TO NONEQUILIBRIUM STEADY STATES

Denis J. Evans[1,2], Stephen R. Williams[1], Debra J. Searles[3,4] and Lamberto Rondoni[5,6]

[1]Research School of Chemistry, Australian National University, Canberra, ACT 0200, Australia

[2]Department of Applied Mathematics, Research School of Physics and Engineering, Australian National University, Canberra, ACT 0200, Australia

[3]Australian Institute for Bioengineering & Nanotechnology Centre for Theoretical and Computational Molecular Science, The University of Queensland, St Lucia, QLD 4072, Australia

[4]School of Chemistry and Molecular Biosciences, The University of Queensland, St Lucia, QLD 4072, Australia

[5]Dipartimento di Scienze Matematiche and Graphene@Polito Lab, Politecnico di Torino, Corso Duca degli Abruzzi 24, 10129 Torino, Italy

[6]INFN, Sezione di Torino, Via P. Giuria 1, 10125 Torino, Italy

# ABSTRACT

The issue of relaxation has been addressed in terms of ergodic theory in the past. However, the application of that theory to models of physical interest is problematic, especially when dealing with relaxation to nonequilibrium steady states. Here, we consider the relaxation of classical, thermostatted particle systems to equilibrium as well as to nonequilibrium steady states, using dynamical notions including decay of correlations. We show that the condition known as $\Omega$T-mixing is *necessary* and *sufficient* to prove relaxation *of ensemble averages* to steady state values. We then observe that the condition known as weak T-mixing applied to smooth observables is *sufficient* for relaxation to be independent of the initial ensemble. Lastly, weak T-mixing for integrable functions makes relaxation independent of the ensemble member, apart from a negligible set of members enabling the result to be applied to observations from a single physical experiment. The results also allow us to give a microscopic derivation of Prigogine's principle of minimum entropy production in the linear response regime. The key to deriving these results lies in shifting the discussion from characteristics of dynamical systems, such as those related to metric transitivity, to physical measurements and to the behaviour of observables. This naturally leads to the notion of *physical ergodicity*.

# I. INTRODUCTION

The study of the relaxation of systems made of many microscopic constituents obeying deterministic time-reversible equations of motion to stationary states, is long and celebrated; it finds its modern roots in Ludwig Boltzmann's Kinetic Theory. Boltzmann's mechanical *and* statistical approach was based on an assumption that he called ergodicity, although what he precisely meant by that has been the object of debate till the present day [1, 2]. Despite this debate, in the early XX[th] century the issue of ergodicity was already considered by physicists to be a practically solved issue due to the results of Enrico Fermi who proved a theorem apparently confirming the common opinion that a certain form of ergodicity was to be expected for the Hamiltonian particle systems of physical interest [3]. In practice, for rather general Hamiltonian systems with $n$ degrees of freedom, but under a very restrictive regularity condition, Fermi proved that given *any* two *open* surface elements $\sigma$ and $\sigma''$ of the $2n$-1 dimensional surface of constant energy, there are trajectories beginning in $\sigma$ that pass through $\sigma''$. Subsequently, ergodic theory became by and large a mathematical subject concerning the properties of generic dynamical systems, which include arbitrarily abstract spaces and deterministic evolutions.

The mathematical notion of ergodicity that is commonly used today is called metric transitivity, and in Hamiltonian dynamics (which preserve phase space volumes) it amounts to state that almost all trajectories densely explore the phase space, $\mathcal{M}$ say; the trajectories that do not explore densely $\mathcal{M}$ constitute a set of vanishing phase space volume. This implies that time averages of observables are equal to appropriate equilibrium phase space averages.

This kind of ergodic theory has been the source of many insights of physical interest, but has often been developed with very little reference to the fundamental ingredients of statistical physics, such as the incredibly large numbers of microscopic degrees of freedom, the special properties of the relevant functions of phase, and the importance of relevant space and time scales. Notable exceptions in this respect include the works by Khinchin, Mazur and Van der Linden, and Lee [4, 5, 6]. Physicists in search of a comprehensive description of nonequilibrium behaviour of interacting many-particle systems, have found ergodic theory problematic. This is demonstrated by the fact that the physicists' treatment of dynamical models of these systems often assume ergodicity as a first step; indeed the equality of time and phase space averages for steady state systems is common [7]. However, almost no system comprised of many interacting particles is actually ergodic in the metrically transitive sense.

The mathematics school under Kolmogorov's lead came to realize that Fermi's hypotheses were practically never satisfied for models of interacting particles. Fermi himself, together with Pasta, Ulam and Tsingou, found that numerical simulations of the dynamics of a chain of nonlinear coupled oscillators, later called an FPU chain, failed to result in equipartition of energy which would be guaranteed if the system was ergodic [8, 9, 10]. This fundamental work for both statistical mechanics and dynamical systems theory, helped inaugurate the era of molecular dynamics as well. Presumably, Fermi's purpose was to use the newly developed computers to validate his theorem, which he knew he had not proven in as general terms as it was desired, but the outcome was that it demonstrated its limitations.

In order to see why the requirements Fermi's theorem are so commonly violated, consider $\sigma'$ to be the region of the phase space $\mathcal{M}$ occupied by all the points of the trajectories with initial condition in $\sigma$ ($\sigma' \supseteq \sigma$ because the initial point belongs to its trajectory). The difficulty is that Fermi's argument requires the border between this region and the rest of the phase space to be smooth, so that $\sigma'$ and its complement are distinct, hence the statistical properties of the trajectories in $\sigma'$ and those of the trajectories in the rest of the phase space, could be clearly distinguished from each other. If this is not the case, then although there are points in any open surface element $\sigma''$ (defined above) occupied by $\sigma'$ there might be other points in $\sigma''$ that did not come from $\sigma$. This can occur when the boundary between regions become fractal, or of a lower dimensionality than the phase space, making $\sigma'$ and its complement too finely and intricately intertwined with each other. It is now known that this often occurs and those trajectories that do not densely explore the phase space may cover a set of positive measure. For the cases considered by Fermi this is related to the topical subject of Arnold diffusion [11] and, in general dynamical systems, to the incredible variety of possible basins of attraction (for instance, the case of riddled attractors [12]).

A time stationary state can be defined as one for which ensemble averaged physical properties do not change with time to within the measureable accuracy. This time stationary state is typically formed by evolution from some initial state for a transient period until the properties are no longer observed to change. If the time stationary state is out of equilibrium, we have a nonequilibrium steady state (NESS) in the physical sense. For such a system, suppose that the initial state is described by a probability distribution function $f_0$ on $\mathcal{M}$, and that the evolution of the physical microstates is represented by equations of motion on $\mathcal{M}$. Then, at variance with the behaviour of averages of physical properties, unless the system is at equilibrium the distribution function evolves from the initial transient $f_0$, taking a different form $f_t$ at every time *t*, and never stopping its evolution. Despite this, the physical properties calculated with $f_t$ converge.[1]

In dynamical systems theory a steady state is associated with a construct that is represented by a time independent probability distribution (an invariant measure, in mathematical terms)[2] that does not need be associated with a continuous probability density function when examined in ostensible phase space. If the dynamics are conservative, and the system is at equilibrium, the ensemble amounts to a probability density function, but for dissipative systems, the ensemble has a singular distribution of phase points in ostensible phase space. In the equilibrium case, the dynamical systems theory terminology refers to invariant measures that have an absolutely continuous $f_0$; in the case of dissipative systems it refers to singular invariant measures. This means that, in this dynamical systems framework, a NESS corresponds to a lower dimensional subset $\mathscr{H}$ of $\mathscr{M}$, which has steady state probability 1 but zero probability with respect to $f_0$, because its phase space volume is zero. Therefore, initial points in $\mathscr{M}$ will not lie on $\mathscr{H}$, apart from those in this invariant set of zero volume, and their time evolution will result in the never ending collapse of the phase space probability distribution towards a singular measure that attributes probability 1 to $\mathscr{H}$. Following Milnor [19], we refer to the invariant set, $\mathscr{H}$, as the NESS attractor.

---

[1] It can be stated that the measure associated with $f_t$ converges to give averages of a specified accuracy, which means that averages of physical properties of the system converge, even though the density is evolving.

[2] An invariant probability distribution will never be attained through time evolution from another probability distribution. It is an idealisation for nonequilibrium steady state systems, but can be represented by a distribution function for equilibrium systems. An invariant measure would give phase space averages of all properties that are time-invariant to all limits of accuracy. If this measure has a density (an equilibrium system), then the density itself does not change with time. A probability distribution in phase space represents a collection of objects in a given state, called "ensemble".

Despite the differences between an invariant measure and the time evolved density $f_t$; for a system that reaches a steady state, at sufficiently long times the averages calculated with the invariant measure and the evolved distribution will be equal to within some accuracy.

These and many other reasons, some of which are illustrated below, make it clear that the evolution of probability densities in phase space is not the proper concept to describe relaxation. As we show, when considering relaxation it is necessary to refer to the physical definition of the steady state, i.e. to the evolution of its physical properties, hence to the evolution of the physically relevant functions of phase. Analogously, we regard a system to be in a NESS when its physical properties are stationary and there is dissipation.

One major concept of ergodic theory, which has been associated with the problem of relaxation, is mixing. To understand the relation between ergodicity and mixing, as well as their connection with relaxation in statistical physics, let us recall a few facts.

*(i) Ergodicity*

Consider the deterministic evolution $S^t : \mathcal{M} \to \mathcal{M}$, with notation meaning that $S^t \mathbf{\Gamma} \in \mathcal{M}$ represents the phase at time *t* along a phase space trajectory starting at $\mathbf{\Gamma} \in \mathcal{M}$. The first important feature of ergodicity is the use of phase space averages to express the infinite time average of an observable $O : \mathcal{M} \to \mathbf{R}$, mathematically represented by,

$$\bar{O}(\mathbf{\Gamma}) = \lim_{t \to \infty} \bar{O}(\mathbf{\Gamma};t) \equiv \lim_{t \to \infty} \frac{1}{t} \int_0^t O(S^s \mathbf{\Gamma}) \, \mathrm{d}s \tag{1}$$

for a system whose initial microstate is represented by $\mathbf{\Gamma} \in \mathcal{M}$. Note that in writing (1) we are assuming the long time limit exists. We are therefore assuming that at long times a state is reached in which the time averages of physical observables are time stationary. In an ergodic system, it is postulated that for almost all initial phases $\Gamma$

$$\overline{O}(\Gamma) = \int_{\mathcal{M}} O(\Gamma) \mathrm{d}\mu(\Gamma) = \langle O \rangle_\mu \qquad (2)$$

where, $\langle \cdot \rangle_\mu$ denotes the phase space average with respect to an appropriate probability distribution $\mu$ on $\mathcal{M}$, and, because of the limit in time the averages $\langle O \rangle_\mu$ are *invariant* under the dynamics $S^t$.

*(ii) Relationship between experimental measurements, time-averages and ensemble averages*

In a single experimental measurement, there is always some degree of time averaging. If the system is time stationary (*i.e.* is at equilibrium or in a NESS), is sufficiently large and the measurement is of an intensive global property (such as the pressure of the system), the measured property, $\overline{O}(\Gamma;t)$, will be indistinguishable from the limiting value given by (1), $\overline{O}(\Gamma)$ for a sufficiently large measurement time, *t*. This is because experimental observations of a system will take a period of time, *t*, and therefore represent the average over microscopic states.[3] The larger the system, the shorter the time averaging period needs to be, and in the thermodynamic limit and for a steady state, $\overline{O}(\Gamma;t) = \overline{O}(\Gamma)$. For small systems, or when a local property is being measured, measurement over an extended period of time will be required to obtain convergence towards $\overline{O}(\Gamma)$. In all cases, if the system is ergodic then $\overline{O}(\Gamma) = \langle O \rangle_\mu$. In experiments it is also common to repeat measurements and report

---

[3] Fermi states: *"Studying the thermodynamical state of a homogeneous fluid of given volume at a given temperature (the pressure is then defined by the equation of state), we observe that there is an infinite number of states of molecular motion that correspond to it. With increasing time, the system exists successively in all these dynamical states that correspond to the given thermodynamical state. From this point of view we may say that a thermodynamical state is the ensemble of all the dynamical states through which, as a result of the molecular motion, the system is rapidly passing"* [E. Fermi, *Thermodynamics*, Dover Publications, New York (1956)]

an average over several measurements. This also allows us to give an estimate of the statistical uncertainty in that average. If the system is time stationary but not ergodic, this result might, or might not, give $\langle O \rangle_\mu$, depending on how the system is prepared.

This picture is consistent with the thermodynamic description, which makes sense only within observation times that are neither too short nor too long: times which are vastly separated from both the microscopic and the geological or the astronomical time scales. Furthermore, the thermodynamic description only refers to systems with a number of degrees of freedom of the order of Avogadro's number. Equation (2) can of course be considered beyond these bounds, *e.g.* for nanoscale systems, investigation of local properties, fluctuations and in abstract mathematical investigations. However, one should not be surprised if the thermodynamic features are not reproduced in such frameworks.

*(iii) Ergodicity and mixing*

One aspect of metric transitivity is the validity of equation (2) for almost all initial conditions $\Gamma \in \mathcal{M}$. However an accurate description of the relaxation of a system of interacting particles requires more than just dense exploration of phase space.

Consider, for instance, a popular pedagogical model in ergodic theory, whose phase space is the unit square $\mathcal{M} = [0,1] \times [0,1]$ with periodic boundary conditions. Given $\Gamma = \begin{pmatrix} x \\ y \end{pmatrix} \in \mathcal{M}$, let $v_x$ and $v_y$ be fixed real numbers and let the dynamics be defined by $S^t \begin{pmatrix} x \\ y \end{pmatrix} = \begin{pmatrix} x + v_x t \\ y + v_y t \end{pmatrix}$ modulo reinjection at the sides of the square. In this case, all trajectories move at constant *phase space* velocity $v = \begin{pmatrix} v_x \\ v_y \end{pmatrix}$ and for irrational $v_x/v_y$ they explore $\mathcal{M}$ densely. Consequently, the only probability density that is invariant under $S^t$ is the uniform distribution in $\mathcal{M}$, $d\mu = dxdy$. Furthermore, no other initial distribution will relax towards the uniform distribution. Figure 1a), illustrates this point, showing that a circular set of phase points never spreads to uniformly cover the square, even though each trajectory visits all regions of $\mathcal{M}$.[4]

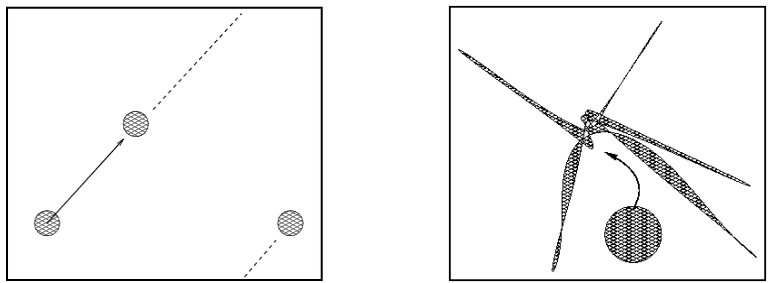

**Figure 1**: Left panel: a circular set of phase space points moves keeping its shape and size. If the ratio of the two components of the phase space velocity is irrational, each point explores densely $\mathcal{M}$, and time averages are reproduced integrating with respect to the uniform phase space distribution. However, points do not spread to uniformly cover $\mathcal{M}$. Right panel: $\mathcal{M}$ is more and more uniformly covered by an evolution which stretches the initial set of points in thinner and thinner filaments, keeping their initial area. Also in this case trajectories explore densely the phase space and time averages are reproduced by integration with respect to the uniform distribution in $\mathcal{M}$, but in a very different fashion.

Figure 1b) shows an alternative way, called *mixing,* in which the points in phase space can evolve ergodically: in this case the distance between phase space points changes with time and might eventually be distributed throughout the accessible space. The mixing behaviour described here is found in many examples of non-dissipative chaotic dynamics. If phase points behaved in phase space like the molecules of a gas behave in real space, one would conclude that ergodicity alone cannot be used to explain relaxation, and that one should invoke the *mixing* condition which implies and is actually stronger than ergodicity. However, we note that trajectories of interacting molecules are very different to phase space trajectories; for example, the phase space points do not interact; so this conclusion cannot be made based on that argument.

---

[4] This dynamics, if intended for real velocities in real space, would also represent that of a set of non-interacting particles all with the same constant velocity moving in a periodic, two-dimensional unit cell. Like the phase space dynamics, for irrational $v_x/v_y$, each particle could access all positions in the unit cell given sufficient time, however the minimum image distance between particles would remain the same at all times just as close points in the phase space $\mathcal{M}$ remain close all for all time (modulo periodic boundaries). This real space dynamics is a very poor model of the physics of the molecules of a gas in a closed container which would have a distribution of velocities, interactions between particles and with the walls resulting in changes in the distance between them etc. The phase space picture that represents a single particle is even poorer, since a single particle cannot relax to a uniform mass distribution.

There are two equivalent formulations of mixing which, like ergodicity, refer to an invariant measure $\mu$, and which are not restricted to conservative dynamics. Given two ($\mu$- measurable) sets $E$ and $F$ in $\mathcal{M}$, mixing is verified if:

$$\left[\mu(E \cap S^t F) - \mu(E)\mu(F)\right] = \left[\mu(S^{-t} E \cap F) - \mu(E)\mu(F)\right] \xrightarrow[t \to \infty]{} 0 \quad (3)$$

where for a generic measurable set $D \subset \mathcal{M}$, $\mu(D) = \int_D d\mu$ is the probability of $D$, and $\mu(\mathcal{M}) = \int_\mathcal{M} d\mu = 1$ by definition of probability measure. If $\mu$ has a density $f$ (*i.e.* it is absolutely continuous with respect to the phase space volumes), one may write $\mu(D) = \int_D f(\Gamma) d\Gamma$. Equation (3) means that the fraction of the set that was initially in $F$ and is then found in any other set $E$ at large times, equals the probability of observing the set $F$ in $\mathcal{M}$. This is possible if the evolving set $S^t F$ spreads all over $\mathcal{M}$, or over the region of $\mathcal{M}$ concerning the steady state, cf. Figure 1b. Equation (3) is equivalent to the following decay of correlations for all integrable observables $A$ and $B$:

$$\left[\int_{\mathcal{M}} A(S^t\mathbf{\Gamma})B(\mathbf{\Gamma})\mathrm{d}\mu(\mathbf{\Gamma}) - \int_{\mathcal{M}} A(\mathbf{\Gamma})\mathrm{d}\mu(\mathbf{\Gamma})\int_{\mathcal{M}} B(\mathbf{\Gamma})\mathrm{d}\mu(\mathbf{\Gamma})\right]$$

(4)

$$= \left[\left\langle \left(A \circ S^t\right)B\right\rangle_\mu - \left\langle A\right\rangle_\mu \left\langle B\right\rangle_\mu\right] \equiv C_\mu(t) \xrightarrow[t \to \infty]{} 0$$

where $A \circ S^t$ represents the composition of the function $A$ with the time evolution, *i.e.* $A \circ S^t\mathbf{\Gamma} \equiv A(S^t\mathbf{\Gamma}) \equiv A(\mathbf{\Gamma}(t))$, and we have defined the correlation function $C_\mu$. The equivalence of (3) and (4) is obtained noting $\int_E \mathrm{d}\mu = \int_{\mathcal{M}} \chi(\mathbf{\Gamma})\mathrm{d}\mu(\mathbf{\Gamma})$.[5]

This result has been be used to show that averages over an appropriate initial distribution approach those taken over the invariant mixing measure that is preserved by the dynamics [12, 13] (see Ref. [14] for a critical analysis). For instance, consider a dynamical system on the unit square $\mathcal{M} = [0,1] \times [0,1]$, which is mixing for the microcanonincal measure $\mathrm{d}\mu_{mc} = \mathrm{d}x\mathrm{d}y = \mathrm{d}\mathbf{\Gamma}$, where $\int_{\mathcal{M}} \mathrm{d}\mu_{mc} = \int_{[0,1]\times[0,1]} \mathrm{d}\mathbf{\Gamma} = 1$ (so one does not need to keep track of the normalization $1/\int_{\mathcal{M}} \mathrm{d}\mathbf{\Gamma}$ of $\mu_{mc}$). Let the initial phase space distribution of the system be given by a density $f_0(\mathbf{\Gamma})$, which is not necessarily uniform but is normalized by definition of probability, $\int_{\mathcal{M}} f_0(\mathbf{\Gamma})\mathrm{d}\mathbf{\Gamma} = 1$, and let the distribution at time $t$ be given by $f_t(\Gamma)$, where $\mathrm{d}\mu_t(x,y) = f_t(x,y)\mathrm{d}x\mathrm{d}y$. One can then write:

$$\langle A \rangle_t = \int_{\mathcal{M}} A(\Gamma) f_t(\Gamma) \mathrm{d}\Gamma$$

$$= \int_{\mathcal{M}} A(S^t\Gamma) f_0(\Gamma) \mathrm{d}\Gamma \qquad (5)$$

$$= \langle (A \circ S^t) f_0 \rangle_{\mu_{mc}}$$

where the second equality is due to the equivalence of the Heisenberg and Schrödinger representations of phase space averages (for a discussion see [7]) and $\langle \cdot \rangle_t$ denotes the average with respect to $\mu_t$. Then using the mixing condition (4),

$$\lim_{t \to \infty} \langle (A \circ S^t) f_0 \rangle_{\mu_{mc}} = \langle A \rangle_{\mu_{mc}} \langle f_0 \rangle_{\mu_{mc}} = \langle A \rangle_{\mu_{mc}} \qquad (6)$$

where the final equality is true because $\langle f_0 \rangle_{\mu_{mc}} = \int f_0 \, \mathrm{d}x \, \mathrm{d}y = 1$. So for systems that are mixing with respect to $\mu_{mc}$ (which implies their dynamics preserves $\mu_{mc}$):

$$\lim_{t \to \infty} \langle A \rangle_t = \langle A \rangle_{\mu_{mc}}.$$

---

[5] The equivalence of condition (3) for all measurable sets and condition (4) for all integrable functions can be understood as follows. In the first place, both Eq. (3) and Eq. (4) imply a loss of memory about the initial conditions: Eq.(3) says that the points found in $E$ at time zero could have come from everywhere else in the phase space; Eq. (4) says that the correlation between any two observables is lost in time. Secondly, observe that $\mu(E)$ equals $\int_E \mathrm{d}\mu = \int_{\mathcal{M}} \chi(\mathbf{\Gamma})\mathrm{d}\mu(\mathbf{\Gamma})$, where the characteristic function $\chi$ is defined by $\chi(\mathbf{\Gamma}) = 1$ if $\mathbf{\Gamma} \in E$ and $\chi(\mathbf{\Gamma}) = 0$ if $\mathbf{\Gamma} \notin E$. It follows that Eq. (3) holds if correlations between integrable functions, which include $\chi$, decay. At the same time, linear combinations of characteristic functions approximate to arbitrary precision integrable functions, therefore the validity of Eq. (3) can be used to imply Eq. (4) [I.P. Cornfeld, S.V. Fomin, Ya. G. Sinai, *Ergodic Theory*, Springer-Verlag, New York (1982)].

The conclusion is that for such systems $\mu_{mc}$ ensemble averages of observables $A$ converge in time from their initial values $\langle A \rangle_0$ to the asymptotic values $\langle A \rangle_{\mu_{mc}}$ corresponding to the uniform distribution in phase space. This result has been known for some time [13, 14]. Umbrella sampling can be used to extend this argument to other ensembles [15].

Because mixing implies ergodicity, if the ensemble averages in the long time limit converge, then time averages commencing from a point on $\mathcal{M}$ will also converge to this result and we can say that the system relaxes. However, mixing refers to the invariant steady state measure hence, in addition to the difficulties mentioned above, it cannot describe the physically relevant transient period. In this paper we determine the conditions required for relaxation to steady states for such systems, and will identify conditions that lead to relaxation in a single experiment.

We show that the condition known as $\Omega$T-mixing is necessary and sufficient to prove relaxation of the ensemble averages of properties to steady state values, while the condition known as weak T-mixing (wT-mixing) applied to smooth observables is sufficient for relaxation to be independent of the initial ensemble. In turn, wT-mixing for integrable functions makes relaxation independent of the initial phase point, apart from a negligible set of phase points, in accord with what is observed experimentally, e.g. in thermodynamic systems. This is achieved by shifting the discussion from properties of dynamical systems such as metric transitivity, to physical measurements and to the behaviour of observables, and leads to the notion of *physical ergodicity*.

## II. THE DISSIPATION FUNCTION AND ΩT-MIXING

In the present paper we firstly use a mathematical notion recently introduced and known as ΩT-mixing, in order to study the ensemble relaxation towards NESSs. Application of similar arguments to relaxation to equilibrium has been given in [14, 16, 17], and relaxation to NESSs using this approach was discussed in [18]. Like equilibrium states, NESSs have stationary, time independent averages of suitably smooth phase functions (*e.g.* pressure, stress, energy *etc.*). However unlike equilibrium states, in dynamical systems theory the NESS of the dissipative dynamics is characterized by an idealized invariant measure that attribute positive probability to sets of dimension lower than the ostensible dimension of phase space. [6] With dissipative dynamics, the phase space $\mathcal{M}$ may contain more than one NESS attractor (*i.e.* in Milnor's sense [19]), which may coexist, since they occupy only a vanishing volume in $\mathcal{M}$.

We will see that in order to investigate the relaxation problem both at and away from equilibrium, one can rely on a quantity introduced in 2000, namely the Dissipation Function $\Omega$ [20, 21]. Given an autonomous dynamical system $\dot{\mathbf{\Gamma}} = G(\mathbf{\Gamma})$ and an initial distribution $f_0$ on $\mathcal{M}$, the dissipation function integrated over the time interval $[0, t]$ is a phase variable defined by:

$$\Omega_{0,t}^{(f_0)}(\mathbf{\Gamma}) = \int_0^t \Omega^{(f_0)}(S^u \mathbf{\Gamma}) du = \ln \frac{f_0(\mathbf{\Gamma})}{f_0(\mathbf{M}^T S^t \mathbf{\Gamma})} - \Lambda_{0,t}(\mathbf{\Gamma}) \qquad (7)$$

---

[6] As a consequence of this lower dimensionality, functions of the phase space probability density, such as the Gibbs entropy, are not defined in a NESS. Alternatively, if followed in its evolution under dissipative dynamics from an initial equilibrium state, towards the NESS, the Gibbs entropy diverges at a constant average rate towards negative infinity [7]. The Gibbs entropy ceases to be a useful concept for nonequilibrium systems in general, and nonequilibrium steady states in particular. This can also be seen as a consequence of the divergence to positive infinity of the probability density, observed from almost any point moving in $\mathcal{M}$, combined with the shrinking volume of the occupied phase space.

where $M^T$ is the time reversal map, *e.g.* $M^T\Gamma \equiv (\mathbf{q}_1,...\mathbf{q}_N,-\mathbf{p}_1,...,-\mathbf{p}_N)$ for systems of $N$ point particles, $\Lambda = \text{div } G$ is the phase space volume variation rate, and for every observable $A$ the subscripts $0,t$ denote integration from time 0 to time $t$ along the trajectory passing in $\Gamma$ at time 0:

$$A_{0,t}(\Gamma) = \int_0^t A(S^u\Gamma) du . \tag{8}$$

Typically, the distribution $f$ is not invariant and changes in time under the dynamics. This does not prevent $\Omega$ from being a phase variable, because equation (7) only refers to the distribution function at a single time.

Definition (7) requires that for all $\Gamma$ where $f_0(\Gamma) \neq 0$, $f_0(M^T S^t \Gamma) \neq 0$, a condition referred to as *ergodic consistency*. Ergodic consistency guarantees (except possibly for a set of measure zero) that the probability density at one point of a Loschmidt's trajectory/antitrajectory conjugate pair is positive if the corresponding point has positive probability density.

The dissipation function depends on the initial distribution, as stressed by the notation in equation (7). However, for sake of simplicity, we will omit the superscript $f_0$ when there is no danger of confusion.

Dividing $\Omega_{0,t}(\Gamma)$ by $t$ and relying on the continuity in time of $f_0(S^t\Gamma)$ and of $\Lambda(S^t\Gamma)$, the instantaneous value $\Omega$ of the dissipation function is obtained from its integral representation (7) (see e.g. [16]).

In the case that $f_0$ is the canonical equilibrium distribution corresponding to no driving, $\Omega$ represents the energy dissipation, a quantity that can be computed or measured in experimental systems, regardless of how near or far the system is from equilibrium [14, 22-24]. If $S^t$ and $f_0$ are time reversal invariant, i.e. $M^T S^t = S^{-t} M^T$

and $f_0(M^T\Gamma) = f_0(\Gamma)$, as appropriate for an equilibrium probability density, $\Omega$ is odd with respect to time reversal, $\Omega(M^T\Gamma) = -\Omega(\Gamma)$, as appropriate for dissipation. Consequently, its average with respect to any distribution function $f$ that is time reversal invariant vanishes: $\langle\Omega\rangle_f = 0$. We make use of this property to introduce the condition called *ΩT-mixing*, i.e. the condition that the following limit results in a finite real number, $L_A$:

$$\lim_{t\to\infty} \int_0^t \langle (A \circ S^s)\Omega\rangle_0 \, ds = L_A \in \mathbf{R} . \tag{9}$$

Because $\langle\Omega\rangle_0 = 0$, $\langle(A \circ S^t)\Omega\rangle_0$ is equal to the correlation function $C_0(t) = \langle(A \circ S^t)\Omega\rangle_0 - \langle A \circ S^t\rangle_0 \langle\Omega\rangle_0$ which is required by (9) to vanish faster $1/t$. This condition is particularly useful in connection with the Dissipation Theorem [25] for the response of a phase variable $A$, which in general terms reads:

$$\langle A\rangle_t = \langle A \circ S^t\rangle_0 = \langle A\rangle_0 + \int_0^t \langle(A \circ S^s)\cdot\Omega\rangle_0 \, ds . \tag{10}$$

### II.1 An example: the isokinetic particle system

For concreteness, let us discuss these issues in the context of a nonequilibrium molecular dynamics model: a system of $N$ particles subject to the following equations of motion:

$$\dot{\mathbf{q}}_i = \mathbf{p}_i/m + C_i\mathbf{F}_e, \quad \dot{\mathbf{p}}_i = \mathbf{F}_i + D_i\mathbf{F}_e - S_i\alpha_{IK}\mathbf{p}_i + S_i\mathbf{F}_{th} \tag{11}$$

In these equations $\mathbf{F}_e$ is an external dissipative field (*e.g.* an electric field applied to a molten salt), and the scalars $C_i$ and $D_i$ couple the system to the field. The system can easily be generalized to tensor coupling parameters if required. If we denote a set of $N_{th}$ thermostatting particles as belonging to the set *th*, $S_i$ is a switch to determine whether particle *i* is a member of the set ($S_i = 0, i \notin th$, $S_i = 1, i \in th$), the thermostat multiplier [7] $\alpha_{IK}$ is chosen to fix the kinetic energy of the thermostatting particles at the value $K_{th}$ and $\mathbf{F}_{th}$ is a fluctuating force to fix the momentum of the thermostatting particles, which is selected to have a value of zero, $\mathbf{P}_{th} = \sum_{i=1}^{N} S_i \mathbf{p}_i = \mathbf{0}$. We assume the interatomic forces $\mathbf{F}_i$, $i = 1,...,N$, are smooth and short ranged functions of the interparticle separation. We also assume that in the absence of the thermostatting and momentum zeroing forces, the equations of motion preserve phase space volumes (*i.e.* $\partial/\partial\mathbf{\Gamma} \cdot \dot{\mathbf{\Gamma}}^{ad} \equiv \Lambda^{ad}(\mathbf{\Gamma}) = 0$) where $\mathbf{\Gamma} \equiv (\mathbf{q}_1,...\mathbf{p}_N)$ is the phase space vector and *ad*, an abbreviation for adiabatic, denotes the fact that the time derivative is calculated with the thermostatting and momentum zeroing forces turned off. This condition is known as the adiabatic incompressibility of phase space condition, or AI$\mathbf{\Gamma}$ [7].

We assume the system of particles is subject to infinite checkerboard boundary conditions [7] – at least in the direction of the force. This means that angular momentum is not a constant of the motion. It also means that dissipation can go on forever without the system relaxing to equilibrium. Currents can flow in the direction of the force forever. The thermostatting particles may be taken to be solid particles, like the walls parallel to the field, which can absorb or liberate heat that may be required to generate a NESS characterized by a fixed value for the kinetic energy of the thermostatting particles.

As observed above, in finite, autonomous, mixing Hamiltonian systems, ensemble averages of smooth phase variables eventually relax to microcanonical averages (also see [14]). If these same systems are thermostatted as in equation (11), there is no field applied and the dynamics is mixing with respect to the canonical distribution, the same argument given in equations (5)-(6) for the microcanonical state can be repeated for the canonical equilibrium (see [15]).

In our system the application of infinite checkerboard boundary conditions means that space is translationally homogeneous but orientationally anisotropic. There are no walls with normals parallel to the field to stop particle currents. The flow (at least parallel to the field) can continue forever.

As the initial equilibrium distribution, we select the distribution that is invariant for the system (11) with vanishing $\mathbf{F}_e$. This is a canonical density augmented with the necessary delta functions, referred to as the isokinetic canonical distribution:

$$f_0(\mathbf{\Gamma}) = \frac{\exp[-\beta_{th} H_0(\mathbf{\Gamma})]\delta(\mathbf{P}_{th})\delta(K_{th}(\mathbf{\Gamma}) - K_{\beta,th})}{\int \exp[-\beta_{th} H_0(\mathbf{\Gamma})]\delta(\mathbf{P}_{th})\delta(K_{th}(\mathbf{\Gamma}) - K_{\beta,th}) d\mathbf{\Gamma}} \qquad (12)$$

where $\mathbf{P}_{th} = \sum_{i=1}^{N} S_i \mathbf{p}_i$ is the total momentum of the thermostatting particles, $K_{th}(\mathbf{\Gamma}) = \sum S_i p_i^2 / 2m_i$ is the kinetic energy of the thermostatting particles and $K_{\beta,th} = (3N_{th} - 4)/(2\beta_{th})$ is the fixed *value* of the kinetic energy of the thermostatting particles. The number of particles in a unit cell is $N$. The kinetic energy of the thermostatting particles is fixed using the Gaussian multiplier $\alpha_{IK}$,

$$\alpha_{IK} = \frac{\sum_i S_i(\mathbf{F}_i + D_i \mathbf{F}_e + S_i \mathbf{F}_{th}) \cdot \mathbf{p}_i}{\sum_i S_i \mathbf{p}_i \cdot \mathbf{p}_i}, \qquad (13)$$

in the equations of motion. Here $\beta_{th} = 1/k_B T_{th}$ where $k_B$ is Boltzmann's constant and for isokinetic systems $T_{th}$ is the kinetic temperature of the thermostatting particles. It is also the equilibrium thermodynamic temperature the system will relax to if it is so allowed. Because the total momentum of the system averages to zero, the equilibrium internal energy of the N-particles in the unit cell is the average of $H_0(\Gamma) = K(p) + \Phi(q)$ under the distribution $f_0$, $\langle H_0 \rangle = \int H_0(\Gamma) f_0(\Gamma) d\Gamma$, where $K$ and $\Phi$ are respectively the kinetic and potential energy of all the particles in the original unit cell. For any particle in the original unit cell and at any time, the potential energy may very well involve interactions with particles that were not, or are not, located in the original unit cell.

We should now specify the ostensible phase space domain that is not referred to explicitly in (12). In the full canonical ensemble the particle momenta are unbounded, however the delta functions in the isokinetic canonical ensemble place four constraints on the momenta of some of the particles in the system so this is no longer the case. The initial coordinates of the particles will each be within some finite range, $\pm L$, within the unit cell of the periodic system. Due to the periodicity, any particle and its environment is essentially identical to any periodic image of that particle. Particles can always be "re-imaged" back into the original unit cell. However calculating certain quantities may have spurious discontinuities if this is done. Thermodynamic quantities like pressure, internal energy etc. are all continuous in time, independent of whether particles are "imaged" in the unit cell.

The thermostatting region that is unnatural can be made arbitrarily remote from the natural system of interest. The thermostatting particles may be buried far inside realistic walls that contain the nonequilibrium flow. In that case it means there is no way that the particles in the system of interest can "know" how heat is ultimately being removed from the system. The thermostats are important as a book keeping device to track the evolution of phase space volumes in a deterministic but open system.

A key point in the definition of the dissipation function, equation (7), is that $\mathbf{\Gamma}$ and $M^T S^t \mathbf{\Gamma}$ are the initial phase points for a trajectory and its conjugate (antitrajectory) respectively. This places constraints on the propagator, $S^t$. For a system defined by equation (11), satisfying the AI$\mathbf{\Gamma}$ condition and that is initially in equilibrium with distribution function (12), it is easy to show that $\Omega$ can be written as:

$$\Omega(\mathbf{\Gamma}) = -\beta_{th} \mathbf{J}(\mathbf{\Gamma}) V \cdot \mathbf{F}_e \tag{14}$$

where $V$ is the volume of the unit cell of our infinitely periodic system and the dissipative flux [7] is given by:

$$\sum_{i=1}^{N} [\frac{\mathbf{p}_i}{m} D_i - \mathbf{F}_i C_i] \cdot \mathbf{F}_e \equiv -\mathbf{J}(\mathbf{\Gamma}) V \cdot \mathbf{F}_e = \dot{H}_0^{ad}. \tag{15}$$

In fact (14, 15) define what we call the *primary* dissipation function for this system. If the field is set to zero there is no dissipation because the initial distribution is the equilibrium distribution for the zero field dynamics (11). In the linear regime, the dissipation function is equal to the so-called entropy production rate.

In the case of the equations of motion (11) and initial distribution (12), the Dissipation Theorem (10) can be written as:

$$\langle A \rangle_{t,\mathbf{F}_e} = \langle A \rangle_{0,0} - \beta_{th} V \int_0^t \langle (A \circ S^s) \mathbf{J} \rangle_{0,\mathbf{F}_e} \cdot \mathbf{F}_e \, ds \tag{16}$$

where the various physical ingredients are explicated, and the notation stresses their roles: $\langle \cdot \rangle_{t,\mathbf{F}_e}$ is the ensemble average with respect to the phase space density $f_t$ which evolves from the initial $f_0$ according to the full field-dependent dynamics, denoted for simplicity by $S^t$ instead of $S^t_{\mathbf{F}_e}$; and therefore $\langle (A \circ S^s) \mathbf{J} \rangle_{0,\mathbf{F}_e}$ means the average with respect to $f_0$ with the evolution of $A$ carried out with the field-driven dynamics.

Expressions (10,16) are exact, arbitrarily near or far from equilibrium and also for systems of arbitrary size. They look similar, but they differ from the linear response expressions for the evolution of phase variables in that the time correlation functions are those determined with the field driven dynamics in (10, 16), whereas the equilibrium time correlation functions appear in linear response theory expressions. Equation (16) shows that if the driving field vanishes, the ensemble averages of phase functions are time independent, provided $f_0$ is invariant for the field-free dynamics. If the system starts with the equilibrium distribution (12), the distribution is preserved by the field free, thermostatted dynamics.

Using the definition (9), the average long time response of $A$ given by (16) yields a real number in the long time limit,

$$\lim_{t \to \infty} \langle A \rangle_{t, \mathbf{F}_e} = \langle A \rangle_{0,0} + L_{A, \mathbf{F}_e} \tag{17}$$

if the system is $\Omega$T-mixing. Because property (9) is necessary and sufficient for this result, another statement of $\Omega$T-mixing in our example is that all phase variables satisfy the condition (17). The limiting value can depend on the value of $\mathbf{F}_e$ and on the number of particles $N$.

## III. CONVERGENCE OF ENSEMBLE AVERAGES OF PROPERTIES UNDER $\Omega$T-MIXING

We would like to find the conditions under which convergence of ensemble averages also correspond to relaxation to a NESS for a single system. To answer this question, let us begin by observing that the NESS attractors for dissipative dynamics concentrate the probability density on sets whose dimension is less than that of the ostensible phase space, and that this dimension decreases as the average dissipation increases [26]. For some dynamics more than one NESS attractor will exist.

Therefore, different systems starting from different phases $\Gamma \in \mathcal{M}$, although pertaining to the same equilibrium state, could evolve towards different asymptotic states yielding different time averages. In that case, because the phase space averages (10, 16, 17) run over all initial phases, $\lim_{t \to \infty} \langle A \rangle_{t, \mathbf{F}_e}$ would be a weighted average of the averages pertaining to the different asymptotic states and, as such, it would not necessarily represent the results of any single experimental measurement. As Fermi's and the more recent works on dynamical systems show, the problem cannot always be cured by separately considering the different basins of attraction in $\mathcal{M}$ because, in general they are too finely intertwined with each other and cannot be separated.

Let us consider the characteristic function $\chi_a^A$ (defined in footnote 5) of the invariant set $E_a^A$ corresponding to a given value $a$ for the time averages of the observable $A$, i.e. let $E_a^A = \{\Gamma \in \mathcal{M} : \overline{A}(\Gamma) = a\}$ be the set of phases $\Gamma$ such that

$$\overline{A}(\Gamma) = \lim_{t \to \infty} \frac{1}{t} \int_0^t A(S^s \Gamma) \mathrm{d}s = a \tag{18}$$

This set is an invariant set because of the limit in $t$, and is disjoint from any other $E_b^A$ with $b \neq a$. Because time averages exist quite generally (Birkhoff-Khinchin Ergodic Theorem [27]), let us assume that the limit in equation (18) exists for all phases except a set of vanishing phase space volume. Then, the union over all $a$, $\bigcup_{a \in \mathbf{R}} E_a^A$, constitutes a set of $\mu_0$-measure 1: $\mu_0 \left( \bigcup_{a \in \mathbf{R}} E_a^A \right) = 1$.

Ensemble averages equal time averages starting from a single $\Gamma$ (or a single experimental determination of the property of the system) if $\mu_0(E_a^A) = 0$ for all values $a$ except one, $\hat{a}$ say, so that $\mu_0(E_{\hat{a}}^A) = 1$. In general, this does not need to be the case and, as usual in response theory, $\Omega$T-mixing constitutes a condition for relaxation on average, rather than starting from a single $\Gamma$. As observed above, $\Omega$T-mixing is *necessary* and *sufficient* for the convergence in the ensemble sense. This result is completely general, because equation (10) and its various versions are exact and directly derived from the dynamics. It does not necessarily concern the results of a measurement on a single experimental systems, but it always concerns ensemble of objects, hence it is appropriate *e.g.* in the case of collections of small systems, or of repetitions of any experiment.

The characteristic function of an invariant set,[7] including those of unions of sets such as $E_a^A$, will obey the $\Omega$T-mixing condition for all dynamics and initial ensembles. For every finite time $t$, equation (10) and the equality for invariant sets, $\mu_t(E) = \mu_0(S^{-t}E) = \mu_0(E)$, yield:

$$\mu_t(E_a^A) = \langle \chi_a^A \rangle_t = \langle \chi_a^A \rangle_0 + \int_0^t \langle (\chi_a^A \circ S^s)\Omega \rangle_0 \, ds = \mu_0(E_a^A) = \langle \chi_a^A \rangle_0 \qquad (19)$$

which means

$$\langle (\chi_a^A \circ S^t)\Omega \rangle_0 = \int \chi_a^A(S^t\Gamma)\Omega(\Gamma) f_0(\Gamma) d\Gamma = \int_{E_a^A} \Omega(\Gamma) f_0(\Gamma) d\Gamma = 0 \qquad (20)$$

for every $t$. Indeed, any constant of the motion will trivially satisfy the condition (17). However, unless all properties satisfy it, the system itself cannot be described as an $\Omega$T-mixing system.

---

[7] A set $E \subset \mathcal{M}$ is called an invariant set for the dynamics $S^t$ if $E = S^t E = S^{-t} E$.

## IV. WEAK T-MIXING

The difficulty of connecting the dynamics of an ensemble of phase points with the dynamics starting from a single point in phase space (or a single experiment) suggests that a satisfactory microscopic approach to the problem of relaxation should be freed from the complications of phase space structures and should focus more on physical measurements. Furthermore, the notion of mixing concerns stationary states, hence the decay of correlations stated in equation (4) cannot describe the memory loss characterizing the evolution of transients from an arbitrary initial state to a final steady state. Equation (4) only describes the loss of correlations among microscopic events within a given steady state. That is important, for instance, for macroscopic measurements and the existence of transport coefficients, but is not suitable for relaxation.

To tackle the problem of relaxation in general, we are going to adopt the perspective recently developed in the study of the steady state fluctuation relations [24], which pointed out a dynamical property, later called wT-mixing. Let $f_0$ be an initial probability density in the phase space $\mathcal{M}$. The dynamics $S^t$ is called wT-mixing with respect to $d\mu_0 = f_0 d\Gamma$ if

$$\left[ \left\langle (A \circ S^t) B \right\rangle_0 - \left\langle A \circ S^t \right\rangle_0 \left\langle B \right\rangle_0 \right] = \left[ \left\langle (A \circ S^t) B \right\rangle_0 - \left\langle A \right\rangle_t \left\langle B \right\rangle_0 \right]$$

$$\equiv C_0(t) \xrightarrow[t \to \infty]{} 0$$

(21)

where we used the identity $\left\langle A \circ S^t \right\rangle_0 = \left\langle A \right\rangle_t$, which is a consequence of the conservation of probability, $\mu_t(E) = \mu_0(S^{-t}E)$. The second equality of equation (21) introduces the correlation function $C_0(t)$ of A and B, with respect to the initial distribution $f_0$; its decay in time is due to the decay of the correlations between the initial and the evolving probability distributions, $d\mu_0 = f_0 d\Gamma$ and $d\mu_t = f_t d\Gamma$.

Now, let both $A$ and $B$ be the characteristic function $\chi_E$ of $E$, a set of points in phase space. The validity of equation (21) implies:

$$\left[\mu_0\left(S^{-t}E \cap E\right) - \mu_0\left(S^{-t}E\right)\mu_0(E)\right] = \left[\mu_t\left(E \cap S^tE\right) - \mu_t(E)\mu_0(E)\right] \xrightarrow[t \to \infty]{} 0 \quad (22)$$

If $E$ is an invariant set, its probability does not change in time so $\mu_t(E \cap S^tE) = \mu_t(E) = \mu_0(E)$, and equation (22) shows that $\mu_0(E) = \mu_0(E)^2$, i.e. $\mu_0(E) = 0$ or 1. Clearly this would be inconsistent with the existence of more than one disjoint invariant set of non-zero $\mu_0$-measure. Therefore wT-mixing can be taken to imply that there is only one such set. If the initial distribution was taken as the idealistic invariant steady state distribution, (22) would become the mixing condition and this argument would be the usual argument for an invariant ergodic measure.

However, the argument for the existence of a single steady state made above cannot be directly extended to systems of dissipative interacting particles starting from an arbitrary distribution and relaxing towards a steady state. This argument is not applicable because the dimension of, *e.g.*, the NESS attractor(s) of dissipative dynamical systems is lower than that of the phase space, and their $\mu_0$-measure vanishes. Equation (22) allows for an arbitrary number of invariant sets of zero $\mu_0$-measure. Therefore, equation (22) does not suffice in general to conclude that there is only one steady state.

Nevertheless, a similar argument can be used to demonstrate that in wT-mixing systems, the value of the infinite time average of phase variables is independent of the initial point in phase space. wT-mixing therefore affords a fresh perspective on ergodic notions in physics, because it has been developed while reconciling certain aspects of the dynamics of physical systems, such as the validity of fluctuation relations, with dynamical systems theory, cf. [23, 24]. Moreover, while the standard ergodic notions based on invariant measures are so problematic in relation to the issues of relaxation to NESSs, wT-mixing seems particularly suited for that, since it deals with the decay of correlations with respect to the initial state.

More precisely, the only formal difference between equation (4) (which defines mixing) and equation (21) is that equation (4) refers to an invariant measure, while the initial distribution of equation (21) is not invariant under nonequilibrium dynamics. This formally minor difference leads to a major conceptual difference between the notions of mixing and wT-mixing. While mixing strictly speaks only of the decay of correlations between events within a given steady state, wT-mixing speaks of the loss of correlations between the initial and the evolving probability distributions. In cases in which these distributions characterize the macroscopic states of a given object, this affords a description of the relaxation process.

Consider a system where $E_b^A$ is the set of all points in phase space that have an infinite time average of $A$ equal to $b$: $\overline{A}(\boldsymbol{\Gamma}) = b$. As discussed below equation (18), because of the infinite time average, this is a time-invariant set. Suppose that condition (22) holds for sets such as $E_{a,\delta}^A = \bigcup_{b \in [a, a+\delta)} E_b^A$, for $\delta > 0$. Sets such as $E_{a,\delta}^A$ are invariant sets because they are a union of invariant sets. Therefore, given two different values $a$ and $b$, equation (22) implies that $\mu_0\left(E_{a,\delta}^A\right)$ and $\mu_0\left(E_{b,\delta}^A\right)$ equal 0 or 1, and one of them at most can be 1, if $|b - a| > \delta$. In addition, for any $\delta > 0$, the real numbers

**R** can be expressed as a countable union of disjoint sets like $[a, a+\delta)$, e.g.

$$\mathbf{R} = \bigcup_{n \in \mathbf{Z}} [n\delta, (n+1)\delta) \quad (23)$$

and the measure of the union of all sets is $\mu_0 \left( \bigcup_{n \in \mathbf{Z}} E^A_{[n\delta,(n+1)\delta)} \right) = 1$. Therefore, whatever accuracy $\delta > 0$ we choose, one of the invariant sets must have a $\mu_0$-measure one and all the rest must be of $\mu_0$-measure zero. Therefore, if the system is wT-mixing, for any $\delta > 0$ there is a single value $a \in \mathbf{R}$ for which $\mu_0(E^A_{a,\delta}) = 1$, and $\mu_0(E^A_{b,\delta}) = 0$ for all $b$, $|b - a| > \delta$.

In other words, wT-mixing for invariant sets like $E^A_{a,\delta}$ implies that all single systems, apart from a set of vanishing phase space volume,[8] eventually converge to a state in which the measurements of *A* yield, with arbitrary accuracy, the value *a*.

This does not imply that there is a single NESS attractor: a system with more than one NESS attractor each with $a \leq \overline{A} < a + \delta$, could satisfy wT-mixing if there are no other sets with non-zero $\mu_0$-measure for which $\overline{A} < a$ or $\overline{A} \geq a + \delta$. These intricacies of the phase space structure are irrelevant in our approach, as they should be, because what matters physically are the values of the physical observables.

If the unions of sets $E^O_o$ of another observable, *O* say, also obey equation (22), the same approach can be adopted for the pair *(A,O)*. Because the physically relevant observables required to characterise a physical system are but a few, condition (22) does not appear particularly strong, if the analysis is restricted to them.

The result is not merely an "average" relaxation concerning an ensemble of systems (as usual in response theory and discussed in section II), but it describes the relaxation expected for a single observation of a physical system: all but a negligible

---

[8] We assume $f_0 > 0$ on all $\mathcal{M}$. If this is not the case on a set of positive volume, $\mathcal{M}$ is too large for the initial equilibrium, and we may restrict it to a smaller space.

set of systems relax to the same steady state, as far as measurement of observables can tell.

We use the term *physical ergodicity* to refer to the condition in which time averages, or physical measurements, of a given observable yield the same value.

How common is wT-mixing dynamics? First of all, note that this uniqueness is not the one of standard ergodic theory, because it concerns the initial distribution of phases and not the steady state. This is important, since "*almost all*" points on the NESS attractor of dissipative systems might satisfy the equality of time-averages and ensemble averages over the points on the NESS attractor, however these points have zero $\mu_0$-measure and therefore refer to *"almost none"* in terms of phase space volumes. Moreover, our uniqueness of the time averages, does not even need the uniqueness of the NESS attractor, hence it is weaker than metric transitivity. wT-mixing for invariant sets $E_{a,\delta}^A$ represents a condition that the dynamics must obey to ensure relaxation from almost all initial phases. For Hamiltonian systems, it is not stronger than ergodicity, hence it is weaker than mixing. It can be graded without causing mathematical inconsistencies, by selecting the observables of interest, hence it can be made as weak as needed. Equation (22) for the invariant sets of the observables of interest is also trivially necessary: if $0 < \mu_0\left(E_{a,\delta}^A\right) < 1$, the condition is violated.

**IV.1 Strong ensemble relaxation**

Even if the wT-mixing condition does not hold for all the invariant sets of interest, suppose that it holds for two observables $A$ and $B = h_0/f_0$, where $h_0$ is, for instance, a smooth positive function which vanishes outside a given hypersphere $E \subset \mathcal{M}$ (which is not necessarily an invariant set) of positive radius. Without loss of generality, we may take $\int h_0(\Gamma) d\Gamma = 1$, which makes $h_0$ a smooth probability density supported on $E$. The validity of (21) yields:

$$\left[ \int f_0(\Gamma) B(\Gamma) A(S^t \Gamma) d\Gamma - \int f_0(\Gamma) A(S^t \Gamma) d\Gamma \int f_0(\Gamma) B(\Gamma) d\Gamma \right]$$
$$= \left[ \int h_0(\Gamma) A(S^t \Gamma) d\Gamma - \int f_0(\Gamma) A(S^t \Gamma) d\Gamma \right] = \left[ \langle A \rangle_t^{(h)} - \langle A \rangle_t^{(f)} \right] \xrightarrow[t \to \infty]{} 0 \quad (24)$$

where $\langle A \rangle_t^{(h)}$ and $\langle A \rangle_t^{(f)}$ are the phase space averages of $A$ at time $t$, starting from the ensembles $h_0$ and $f_0$, respectively. Equation (24) states that provided the asymptotic observable value $\langle A \rangle_\mu$ is guaranteed to exist *e.g.* if the system is $\Omega$T-mixing, its value does not depend on the initial ensemble.[9] As a matter of fact, because $f_0$ is an equilibrium distribution, we can assume that it is smooth, hence approximately constant within phase space hyperspheres of sufficiently small radius. Therefore, the condition that $B$ verifies wT-mixing with $A$ (24) can actually be replaced by the condition that $h_0$ does, and $h_0$ can essentially be the characteristic function of a hypersphere of positive radius, as small as one likes, in $\mathcal{M}$. If all hyperspheres verify wT-mixing with $A$, the ensemble averages of $A$ over trajectories starting within all hyperspheres yield the same value. This does not mean that almost all single system time averages converge to that value, as every hypersphere could contain a positive fraction of initial conditions of trajectories producing different time averages. However, it requires the fractions of initial conditions leading to different time averages to be the same everywhere in $\mathcal{M}$, which is peculiar in relation to the phase space description of a physical object and to the measurements which identify its macroscopic state. Again, mixing refers to invariant measures while wT-mixing refers to the known ergodically consistent initial distributions. This has many advantages, including that for our result to hold, it does not matter whether a single steady state attractor is approached in time or not.

---

[9] Recall that $\Omega = \Omega^{(f_0)}$ is defined with respect to $f_0$, hence that $f_0$ is positive over the ostensible phase space $\mathcal{M}$, while $h_0$ does not need to be positive everywhere in $\mathcal{M}$.

## V. PHYSICAL CONSIDERATIONS

Here we summarise the different mixing conditions and their physical implications. $\Omega$T-mixing requires the correlation function with respect to the initial distribution not only to go to zero, but (whether stationary or transient) to vanish sufficiently rapidly for its integral to converge, so that $\lim_{t\to\infty}\langle A\rangle_t$ is *finite*. For example if the equilibrium time correlation function goes as $1/t$ at long times, we will have a logarithmic divergence and the system will not relax to a NESS. This is quite different to the ergodic theory result for autonomous Hamiltonian systems, where mixing implies relaxation on average towards the time independent microcanonical equilibrium distribution, irrespectively of the decay rate of the correlations. However, as pointed out above, this is due to the fact that mixing concerns states with measures that do not evolve. The mixing condition (4-6) cannot be used to prove relaxation from a smooth initial distribution to the invariant NESS distribution because the invariant nonequilibrium distribution is singular [15] and the distribution will never become an invariant distribution: it is evolving at all times.

If we turn briefly to the *transient* time correlation function for the nonlinear response, the mixing condition is simply not relevant. The distributions of states used to compute transient time correlation functions are not stationary.

Equation (10) can be used to derive the Green-Kubo [21, 28, 29] relations in the limit of zero field. However the conditions required are subtle, and different to that used to obtain (16). Kubo's results [28] were for the linearized adiabatic response (*i.e.* no thermostats) of a canonical ensemble of systems. We derived equation (16) for isokinetic dynamics where the kinetic energy of the thermostatting particles is fixed and the distribution for the system of interest is isokinetic canonical – equation (12). Thus the equilibrium time correlation function appearing in (16) is for field free isokinetic dynamics. Therefore to obtain the Green-Kubo relationships, we need to derive the equivalent of (16) using an initial canonical distribution function to generate the initial points, followed by unthermostatted equations of motion to evaluate the correlation function. The resulting equation will look like (16), but as in Kubo's system the time correlation functions will involve canonical distributions but field free, constant energy, Newtonian trajectories. We also note that equation (16) using the isokinetic dynamics and starting from an isokinetic distribution for $\Omega$T-mixing systems are consistent with the result of Evans and Morriss [30] where it was proved that to leading order in the number of degrees of freedom in the system with a correction of order $O(1/N)$, the two equilibrium correlation functions are identical. Of course if the dissipative field only couples to particles in the system of interest and the thermostat region is large and remote, the fluctuations in the dissipation function (which is local to the system of interest) will hardly be affected by the presence or absence of thermostatting terms in the large remote thermostatting region.

Because the thermostat is unphysical, we can make the system more realistic by only thermostatting a remote set of particles. If we only thermostat particles that are remote from the natural system of interest (still within the unit cell), we can always appeal to the gedanken experiment that if we make the thermostatting region ever more remote from the system of interest there is just no way that the physical system of interest can "know" how the remote thermostatting is actually occurring. If the external fields are set to zero and the system is allowed to relax to equilibrium we know the thermodynamic temperature of that underlying equilibrium system. That is the temperature that appears in the equations given above.

There is yet another interesting observation we can make regarding Kubo's system [28, 29]. If you consider viscous flow in a dilute gas then as is known from kinetic theory, the viscosity of a gas increases with temperature. This means that for any finite field, no matter how small, the shear stress of an adiabatic shearing gas must increase with time. This means that a shearing unthermostatted gas can never be ΩT-mixing! In a physical sense for such a system, time correlations either do not decay or do not decay rapidly enough for ΩT-mixing. One can see how this memory effect occurs. If among the initial ensemble members, one encounters a fluctuation that increases the gas viscosity, that fluctuation will, at fixed strain rate, heat the gas slightly. In this slightly heated gas the viscosity will be slightly higher than on average increasing the likelihood of further fluctuations that in turn increase the viscosity. This is a run-away process that prevents the decay of correlations required for the ΩT-mixing condition.

If we assume ΩT-mixing we see that although the long time states predicted by (10, 16, 17) may not be ergodic in the metrically transitive sense, those asymptotic states have nevertheless stationary ensemble averages. If wT-mixing holds for the characteristic functions of the invariant sets of the different values of the observables of interest, the ensemble averages equal the corresponding single system time averages, for almost all initial phases.

# VI. IN THE LINEAR REGIME THE DISSIPATION IS MINIMAL WITH RESPECT TO VARIATIONS OF THE INITIAL DISTRIBUTION

Consider an initial perturbation of the equilibrium canonical distribution of the following form:

$$f_{\lambda g}(\Gamma) = \frac{\exp[-\beta_{th}H_0(\Gamma) - \lambda g(\Gamma)]\delta(\mathbf{P}_{th})\delta(K_{th}(\Gamma) - K_{\beta,th})}{\int \exp[-\beta_{th}H_0(\Gamma) - \lambda g(\Gamma)]\delta(\mathbf{P}_{th})\delta(K_{th}(\Gamma) - K_{\beta,th})d\Gamma} \qquad (25)$$

where the deviation function $g$ is even in the momenta, is nonsingular, real and integrable. The positive number $\lambda$ is a scaling parameter that allows us to scale the magnitude of the deviation from the equilibrium distribution. The dissipation function for this initial distribution and field driven dynamics (11) is

$$\Omega^{(\lambda g)}(\Gamma) = -\beta_{th}\mathbf{J}(\Gamma)V \cdot \mathbf{F}_e - \lambda \dot{g}(\Gamma) \qquad (26)$$

We know that the distribution will not be conserved by the field free dynamics unless $\lambda = 0$, and when $\lambda = 0$ we have the primary dissipation function, $\Omega^{(0)}(\Gamma) = -\beta_{th}\mathbf{J}(\Gamma)V \cdot \mathbf{F}_e$, as defined above.

If the system is $\Omega$T-mixing, time-averages of phase variables will become constant so $\overline{\Omega}^{(\lambda g)} = \lim_{t \to \infty}\langle \Omega^{(\lambda g)}\rangle_{t,\lambda g,\mathbf{F}_e}$ and $\lim_{t \to \infty}\langle \dot{g}\rangle_{t,\lambda g,\mathbf{F}_e} = 0$. If it is also wT-mixing, the long time averages will be independent of the initial perturbation so $\overline{\Omega}^{(\lambda g)} = \lim_{t \to \infty}\langle \Omega^{(\lambda g)}\rangle_{t,\lambda g,\mathbf{F}_e} = \lim_{t \to \infty}\langle \Omega^{(\lambda g)}\rangle_{t,0,\mathbf{F}_e} = \lim_{t \to \infty}\langle \Omega^{(0)}\rangle_{t,0,\mathbf{F}_e} = \overline{\Omega}^{(0)}$.

Furthermore, in the linear regime, the time integral of the primary dissipation function is always less than the time integral of the dissipation with $\lambda \neq 0$. To demonstrate this consider

$$\int_0^t \langle \Omega^{(\lambda g)} - \Omega^{(0)}\rangle_{s,\lambda g} ds = -\lambda \int_0^t \langle \dot{g}\rangle_{s,\lambda g} ds \qquad (27)$$

Now,

$$\langle \dot{g} \rangle_{t,\lambda g, \mathbf{F_e}} = \left( \langle \dot{g} \rangle_{0,\lambda g, \mathbf{F_e}} + \int_0^t \langle (\dot{g} \circ S^s) \Omega_{\lambda g} \rangle_{0,\lambda g, \mathbf{F_e}} \, ds \right)$$

$$= \left( -\beta V \mathbf{F_e} \cdot \int_0^t \langle (\dot{g} \circ S^s) \mathbf{J} \rangle_{0,\lambda g, \mathbf{F_e}} \, ds - \lambda \int_0^t \langle (\dot{g} \circ S^s) \dot{g} \rangle_{0,\lambda g, \mathbf{F_e}} \, ds \right) \quad (28)$$

$$\xrightarrow[\mathbf{F_e} \to 0]{} -\lambda \int_0^t \langle (\dot{g} \circ S^s) \dot{g} \rangle_{0,\lambda g, 0} \, ds = -\lambda \langle \dot{g} \rangle_{t,\lambda g, 0}$$

where we have used $\langle \dot{g} \rangle_{0,\lambda g, \mathbf{F_e}} = 0$ in the first equality. We also know from the second law inequality [25, 31] that for a nonequilibrium system $\int_0^t \langle \Omega^{(\lambda g)} \rangle_{s,\lambda g, \mathbf{F_e}} \, ds > 0$, and therefore when $\mathbf{F}_e = 0$ and there is an initial perturbation away from equilibrium, $-\lambda \int_0^t \langle \dot{g} \rangle_{s,\lambda g, 0} \, ds > 0$. Then by combining (27) and (28), we obtain

$$\lim_{\mathbf{F_e} \to 0} \int_0^t \langle \Omega^{(\lambda g)} - \Omega^{(0)} \rangle_{s,\lambda g, \mathbf{F_e}} \, ds = -\lambda \int_0^t \langle \dot{g} \rangle_{s,\lambda g, 0} \, ds > 0. \quad (29)$$

Therefore,

$$\lim_{\mathbf{F_e} \to 0} \int_0^t \langle \Omega^{(\lambda g)} \rangle_{s,\lambda g, \mathbf{F_e}} \, ds > \lim_{\mathbf{F_e} \to 0} \int_0^t \langle \Omega^{(0)} \rangle_{s,\lambda g, \mathbf{F_e}} \, ds, \quad (30)$$

*i.e.* at sufficiently small fields, and at all times, the time-integrated, ensemble average primary dissipation function is minimal with respect to the initial distribution.

This means that the primary dissipation, which in the long time limit is equal to the steady state dissipation, provides the minimum dissipation for that dynamics at low fields. All forms of dissipation other than the primary dissipation diminish to zero at long times.

This result parallels Prigogine's principle of minimum entropy production in the linear response regime, which concerns the variation in time of the entropy production, but says nothing about its dependence on the initial state [32]:

*In the linear regime, the total entropy production in a system subject to [a] flow of energy and matter, $d_i S / dt = \int \sigma \, dV$, reaches a minimum value at the nonequilibrium stationary state. This is because the unconstrained forces adjust themselves to make their conjugate fluxes go to zero.*"

We have already noted that in the linear regime the average dissipation is equal to the so-called entropy production rate. Therefore, we may also recover Prigogine's principle itself, by making an informed choice of the initial perturbation $g$. In our system there is no net mass flow into or out of the unit cell, therefore we have to construct a second "force" $F_{e,2}$ that is capable of generating a conjugate flux $\dot{g}$. This unconstrained force will adjust itself so that $\dot{g}$ averages to zero in the steady state, as we have seen above. If the equations of motion take the same form as equation (11) but with coupling parameters $C_{2,i}, D_{2,i}$ and a "force" $\mathbf{F}_{2,e}$ we see that we merely have to find the coupling parameters such that $\lambda \dot{g} = \mathbf{F}_{2,e} \cdot \sum \left[ \frac{\mathbf{p}_i}{m} D_{2,i} - \mathbf{F}_i C_{2,i} \right]$. Because of our results above, for wT-mixing systems this is both an ensemble and a single system, or single experiment, result.

## VII. CONCLUSION

We have shown that ΩT-mixing is necessary and sufficient for an initial ensemble to relax to a steady state. wT-mixing, instead, leads to convergence in the sense of time averages of observables, from almost all initial conditions, *i.e.* for almost all single experimental measurements of a system. We refer to this situation as *Physical Ergodicity*, because it is related to measurements of observables, rather than to the properties of the phase space distributions.

These conditions differ substantially from the standard ergodic theory notions, because they refer to the initial probability distribution, and not to invariant measure. Among the numerous consequences of this fact, we have that the relaxation argument expressed by equation (4), makes no reference to the rate at which correlations decay in contrast to that based on ΩT-mixing. This reflects the fact that the argument of equation (4) speaks of the decay of correlations within the NESS attractor, and not of the decay of correlations between an initial and final state.

The condition based on wT-mixing better suits the needs of physical studies of nonequilibrium many-body interacting particles than arguments based on mixing within the NESS attractor because the NESS attractor of this dissipative dynamics occupies zero volume in the ostensible phase space. Therefore even if "almost all" points on the NESS attractor satisfy a given desired property, these points have zero measure in the equilibrium distribution and therefore this means very little for systems starting in a given equilibrium state. Moreover, the wT-mixing condition can be graded to the needs of observations, by restricting it to the variables of physical interest. This frees the dynamics of demanding conditions such as metric transitivity.

If the system is wT-mixing, there may be a set of initial conditions of vanishing $\mu_0$-probability measure that do not yield the same time average, but this is only a set of vanishing phase space volume. Therefore, even in the case of dissipative dynamics the irreversibility of the relaxation process has been connected to "counting" of states, as done in the past for the equilibrium case. In that case, "counting" meant comparing the fractions of phase space pertaining to different states, *e.g.* Ref. [33], and finding that by far the largest fraction is occupied by the equilibrium state [8, 34-36]. In the case of convergence to a NESS of a wT-mixing system, we have shown that by far the largest fraction of the phase space is occupied by phases that yield the same observable value for a given phase variable.

$\Omega$T-mixing, *per se*, implies a weaker result related to relaxation than wT-mixing does. This is, however, a rather strong result which states that relaxation on average does not depend on the initial distribution, if $\Omega$T-mixing holds. In particular, although this is just an ensemble result, the initial ensemble can be as small a set around any $\Gamma \in \mathcal{M}$ as one wishes.

ACKOWLEDGEMENTS

We thank Ms. Charolotte F. Petersen, Dr Owen G. Jepps and Dr James C. Reid for useful comments. We would also like to thank the Australian Research Council for support of this research. LR thanks the European Research Council, for funding under the European Community's Seventh Framework Programme (FP7/2007-2013)/ERC grant agreement n 202680. The EC is not liable for any use that can be made on the information contained herein.

**REFERENCES**

1. G. Gallavotti, *Ergodicity, ensembles, irreversibility in Boltzmann and beyond*, J. Stat. Phys. **78**, 1571 (1994)

2. J. Uffink, *Boltzmann's Work in Statistical Physics*, Stanford Encyclopedia of Philosophy (2014)

3. E. Fermi, *Beweis, daß ein mechanisches normalsystem im algemeinen quasi-ergodisch ist*. Physikalische Zeitschrift **24**, 261 (1923). Reprinted in Fermi, E.: *Note e Memorie (Collected papers)*, Accademia dei Lincei and University of Chicago Press, vol. I (1961)

4. A. Khinchin, *Mathematical Foundations of Statistical Mechanics*, Dover Publications, New York, (1949)

5. P. Mazur, J. van der Linden, *Asymptotic form of the structure function for real systems*, J. Math. Phys. **4** 271 (1963)

6. M.H. Lee, *Ergodic theory, infinite products, and long time behavior in Hermitian models*, Phys. Rev. Lett. **87** 250601 (2001). *Ergodicity in simple and not so simple systems and Kubo's condition*, Physica A **314** 583 (2002). *Why does Boltzmann's ergodic hypothesis work and when does it fail*, Physica A **365** 150 (2006). *Birkhoff's theorem, many-body response functions and the ergodic condition*, Phys. Rev. Lett. **98** 110403 (2007)

7. D.J. Evans, G.P. Morriss, *Statistical Mechanics of Nonequilibrium Liquids*, Cambridge University Press, Cambridge (2008)

8. P. Castiglione, M. Falcioni, A. Lesne, A. Vulpiani, *Chaos and Coarse Graining in Statistical Mechanics*, Cambridge University Press, Cambridge (2008)

9. G. Gallavotti (ed), *The Fermi-Pasta-Ulam Problem: A Status Report*, Springer, Berlin (2008)

10. P. Poggi and S. Ruffo, *Exact solutions in the FPU oscillator chain*, Physica D, **103**, 251-272 (1997)

11. A.J. Lichtenberg, M.A. Lieberman, *Regular and chaotic dynamics*, Springer, New York (1992)

12. E. Ott, *Chaos in Dynamical Systems*, Cambridge University Press, Cambridge (2003)

13. Ya. G. Sinai, *Introduction to Ergodic Theory*, Princeton University Press, Princeton (1976)

14. D.J. Evans, S.R. Williams, L. Rondoni, *A mathematical proof of the zeroth "law" of thermodynamics and the nonlinear Fourier "law" for heat flow*, J. Chem. Phys., **137**, 194109 (2012)

15. D.J. Evans, S.R. Williams, L. Rondoni and D.J. Searles, *Ergodicity of non-Hamiltonian equilibrium systems*, in preparation (2015)

16. D.J. Evans, D.J. Searles and S.R. Williams, *Dissipation and the relaxation to equilibrium*, J. Stat. Mech. P07029 (2009)

17. J.C. Reid, D.J. Evans and D.J. Searles, *Communication: Beyond Boltzmann's H-theorem: Demonstration of the relaxation theorem for a non-monotonic approach to equilibrium*, J. Chem. Phys. **136** 021101 (2012)

18. D.J. Evans, D.J. Searles and S.R. Williams, *On the probability of violations of Fourier's law for heat flow in small systems observed for short times*, J. Chem. Phys. **132** 024501 (2010)

19. J. Milnor, *On the Concept of an Attractor,* Comm. Math. Phys. **99** 177 (1985)

20. D.J. Searles and D.J. Evans, *Ensemble dependence of the transient fluctuation theorem*, J. Chem. Phys. 113 3503 (2000)

21. D.J. Evans and D.J. Searles, *The fluctuation theorem*, Adv. Phys. **52** 1529 (2002)

22. J.C. Reid, S.R. Williams, D.J. Searles, L. Rondoni and D.J. Evans, *Fluctuation Relations and the Foundations of Statistical Thermodynamics:a Deterministic approach and numerical demonstration*, in Nonequilibrium Statistical Physics of Small Systems: Fluctuation Relations and Beyond, Editors R. Klages, W. Just, C. Jarzynski, Wiley-VCH, (2013)

23. D.J. Evans, D.J. Searles and L. Rondoni, *On the application of the Gallavotti-Cohen fluctuation relation to thermostatted steady states near equilibrium*, Phys. Rev. *E* **71** 056120 (2005)

24. D.J. Searles, L. Rondoni, D.J. Evans, *The steady state fluctuation relation for the dissipation function*, J. Stat. Phys. **128** 1337 (2007)

25. D.J. Evans, D.J. Searles and S.R. Williams, *On the fluctuation theorem for the dissipation function and its connection with response theory*, J. Chem. Phys. **128** 014504 (2008). Erratum, *ibid* **128**, 249901 (2008)

26. D.J. Evans, E.G.D. Cohen, D.J. Searles, F. Bonetto, *Note on the Kaplan Yorke Dimension and Linear Transport Coefficients*, J. Stat. Phys, **101**, 17 (2000)

27. I.P. Cornfeld, S.V. Fomin, Ya. G. Sinai, *Ergodic Theory*, Springer Verlag, New York (1982)

28. R. Kubo, *Statistical-Mechanical Theory of Irreversible Processes. I. General Theory and Simple Applications to Magnetic and Conduction Problems*, J. Phys. Soc. Jpn. **12** 570 (1957)

29. M.S. Green, *Markoff Random Processes and the Statistical Mechanics of Time‐Dependent Phenomena. II. Irreversible Processes in Fluids*, J. Chem. Phys. **22** 398 (1954)

30. D.J. Evans and G.P. Morriss, *Equilibrium time correlation functions under gaussian isothermal dynamics*, Chem. Phys. **87** 451 (1984)

31. D.J. Searles and D.J. Evans, *Fluctuation relations for nonequilibrium systems*, Aust. J. Chem. **57** 1119 (2004)

32. I. Prigogine, *Etude Thermodynamique des Processus Irreversibles*, Desoer, Liége (1947); I. Prigogine, *Introduction to Thermodynamics of Irreversible Processes*, John Wiley & Sons, Inc., New York (1967)

33. I. Pitowsky, *Typicality and the Role of the Lebesgue Measure in Statistical Mechanics*, in Y. Ben-Menahem and M. Hemmo (eds.), Probability in Physics, Springer, Berlin (2012)

34. S. Chibbaro, L. Rondoni, A. Vulpiani, *On the Foundations of Statistical Mechanics: Ergodicity, Many Degrees of Freedom and Inference*, Comm. Theor. Phys. **62** 69 (2014)

35. S. Chibbaro, L. Rondoni, A. Vulpiani, *Reductionism, Emergence and Levels of Reality*, Springer, New York (2014)

36. J.L. Lebowitz, *Boltzmann's entropy and time arrow*, Physics Today **46** 32 (1993)